\documentclass[aps,prl,showpacs,showkeys,twocolumn,amssymb,groupedaddress]{revtex4}
\usepackage[colorlinks=true,citecolor=blue,filecolor=blue,linkcolor=blue,urlcolor=blue]{hyperref}
\usepackage{amsmath}
\usepackage{amssymb}
\usepackage{graphicx}

\begin{document}

\title{Light output response of the LVD liquid scintillator to neutron-induced nuclear recoils}

\author{G.Bruno}
\email{gianmarco.bruno@lngs.infn.it}

\affiliation{INFN - Laboratori Nazionali del Gran Sasso
             s.s. 17bis km 18.910, Assergi (AQ), Italy}

\begin{abstract}
The organic liquid scintillator used in the LVD experiment (INFN Gran Sasso National Laboratory) has been exposed to an Am-Be neutron source to measure the light response function for neutron energies in the region from about 4 to 11 MeV.\\
A full Monte Carlo simulation, incorporating the detector response, is used to generate neutron scattering spectra which are matched to the observed ones to determine the quenching factors.\\
The obtained light output response is well described by the semi-empirical Birks model.\\
The results, consistent with those obtained by other authors using similar hydrocarbonate scintillators, can be of interest for fast and high-energy neutron spectroscopy that could be performed with this detector.
\end{abstract}


\maketitle

\section{Introduction}
The Large Volume Detector (LVD), in the INFN Gran Sasso National Laboratory (Italy), below an average 3600 m water equivalent rock overburden, is a 1 kt liquid scintillator detector whose major purpose is monitoring the Galaxy to study neutrino bursts from gravitational stellar collapses \cite{ncim92}.
LVD consists of an array of 840 scintillator counters, 1.5 m$^3$ each, divided in three, identical and independent \emph{towers}. 
Each counter is viewed from the top by three 15 cm photomultiplier tubes (PMTs) \cite{online}.\\
The LVD liquid scintillator is a mixture of aliphatic and aromatic hydrocarbons (C$_{n}$H$_{2n}$ with $\bar{n}$ = 9.6) also known as \emph{White Spirit}.
For historical reasons 73\% of the LVD scintillator was produced in Russia and has an aromatic hydrocarbons content of  $\sim$16$\%$, while the remaining amount has been produced in Italy and contains $\sim$8$\%$ of aromatics.
The scintillator with the higher percentage of aromatics presents a light yield of about 20$\%$ higher. 
We do not expect this slight difference on the scintillator composition to have an impact on the quenching factors therefore we decided to measure the proton light output response in the liquid scintillator produced in Russia that is present in the majority of counters (LVD-scint).\\
In addition, since in 2005 one of the LVD counters (with aromatic hydrocarbons content of $\sim$8$\%$) was doped with gadolinium \cite{gdperf} to improve the signal to noise ratio in neutron detection, we decided to repeat the measurement also on this particular scintillator (LVD-Gd).\\
%
In organic scintillators the light output response, the energy emitted as fluorescence, is non-linear with respect to the energy deposited by ionizing particles (this is particularly true for highly ionizing particles).
This behavior is attributed to quenching of the primary excitation by the high ionization density along the particle track which leads to a decrease of the scintillation efficiency.
In the absence of quenching the light yield is proportional to the energy released, which may be written in differential form as:
\begin{equation}
 \frac{dL}{dx} = S \frac{dE}{dx}
 \label{diffly}
 \end{equation}
where $\frac{dL}{dx}$ is the fluorescent energy emitted per unit path length, $S$ is the scintillation efficiency and $\frac{dE}{dx}$ is the specific stopping power.
To describe the quenching effect, Birks \cite{birks} proposed a semi-empirical relation in which the differential light output is reduced for high energy loss:
\begin{equation}
 \frac{dL}{dx} = \frac{S \frac{dE}{dx}}{1+ kB\frac{dE}{dx}}
 \label{eqbirks}
 \end{equation}
where the product $kB$ is usually treated as a single parameter and it is known as Birks factor.\\
The quenching factor for ions is defined as the ratio of light yield of ions to that of electrons of the same energy. Integrating  eq. \ref{eqbirks} it can be written as:
\begin{equation}
 Q_{i} = \frac{L_{i}(E)}{L_{e}(E)} = \frac{\int_0^E \frac{dE}{1+kB(\frac{dE}{dx})_i}}
                                          {\int_0^E \frac{dE}{1+kB(\frac{dE}{dx})_e}}
 \label{eqq}
\end{equation}

\section{Experimental setup and procedure}
Two scintillator detectors of the same dimensions of the counters of the LVD experiment, have been used to perform the measurements.
Each detector consists of a 1 $\times$ 1 $\times$ 1.5 m$^3$ stainless steel tank filled with 1.2 ton of liquid scintillator and viewed from the top by three PMTs PHOTONIS XP3550 of 5'' diameter.
The signals of each PMT are recorded by a CAEN V1724 waveform digitizer, with 100 MHz of sampling frequency and 137 $\mu V$ of resolution per bit.
The board hosts a Field Programmable Gate Array (FPGA) for each input channel.
The data stream is continuously written in a circular memory buffer, which is frozen by the FPGA when the trigger occurs.
The board implement also the Zero Length Encoding (ZLE) algorithm for data reduction. This algorithm allows to set a threshold, which could be different from the trigger threshold, and discard the data under threshold saving time in both data transfer and processing.
The acquisition window length, the number of samples pre- and post-trigger, the threshold and the level of coincidence among different channels are programmable features.\\
In the center of the detector we have inserted an $^{241}$Am$^{9}$Be source encapsulated in a waterproof housing and held in place by a stainless steel rod.\\
The radionuclide $^{241}$Am decays emitting an alpha particle of 5.6 MeV. Under bombardment by alphas, $^{9}$Be undergoes a nuclear reaction forming the excited compound nucleus $^{13}$C$^{*}$, which then decays mostly producing either $^{12}$C at ground state or at the first excited level \cite{leo}:
\begin{eqnarray}
^{13}C^{*} \rightarrow&^{12}C_{gs}& + \;n \quad 6.5\, \text{MeV} \lesssim E_n \lesssim 11.0      \, \text{MeV}  \label{eqn1}      \\
&&                                                                                                         \nonumber \\
^{13}C^{*} \rightarrow&^{12}C^{*}&  + \;n \qquad\qquad\quad\:\:\:\,   E_n \lesssim 6.5      \:\, \text{MeV} \nonumber \\
                      &^{12}C^{*}& \rightarrow ^{12}C_{gs} + \gamma \quad\:\:\:\,\, E_{\gamma} = 4.44 \, \text{MeV}
\label{eqn2}
\end{eqnarray}
In the latter case the neutron is accompanied by a de-excitation gamma-ray of energy 4.44 MeV.
The source used in these measurements emits about 10 neutrons s$^{-1}$.\\
The emitted neutrons are thermalized by elastic scattering on the hydrogen atoms of the organic scintillator. During the slowing down process, a fraction of the neutron kinetic energy is converted to photons which are detected by the PMTs giving the prompt signal. After being thermalized, neutrons are eventually captured by $^{1}$H, $^{nat}$Gd (when present) and with lower probability by $^{12}$C and $^{56}$Fe nuclei producing the delayed signal.
The mean neutron capture time depends on the target composition (n-capture cross section) and on the detector geometry because neutrons which migrate farther from the source can escape from the detector leading to a shorter mean capture time.
\section{Analysis and Results}
\begin{figure}[htb]
\begin{center}
 \includegraphics[width=20pc,angle=0]{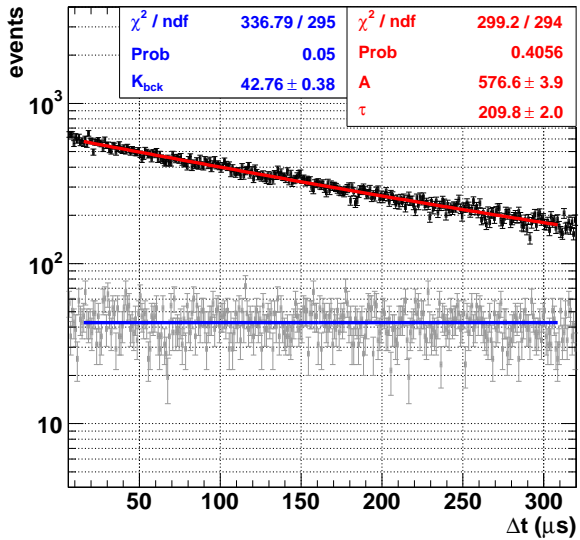}
 \hspace{0.1pc}
 \includegraphics[width=20pc,angle=0]{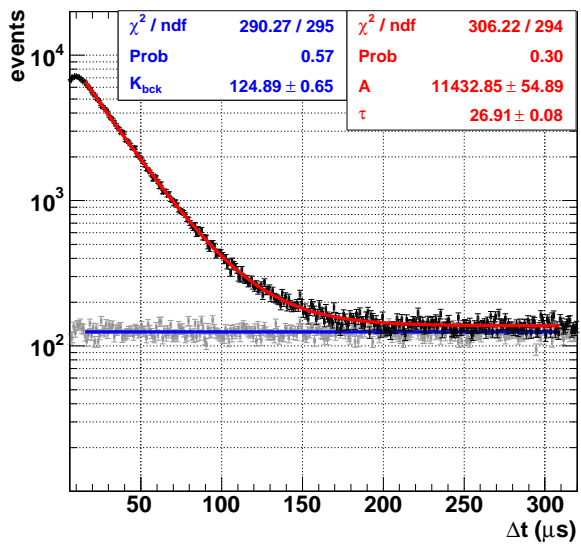}
 \caption{\label{dt} Distribution of the absolute value of the time difference between the trigger and pre- or post-trigger in the same event (blu points) compared with the background level (in gray).
 Shapes are, as expected, flat for the background (fit function: y$=k_{bck}$) and exponential for the data with the neutron source (fit function: y$=Ae^{-t/\tau}+k_{bck}$). Top and bottom panel are relative to LVD-scint and LVD-Gd scintillator respectively.}
 \end{center}
\end{figure}

\begin{figure}[htb]
\begin{center}
 \includegraphics[width=20pc,angle=0]{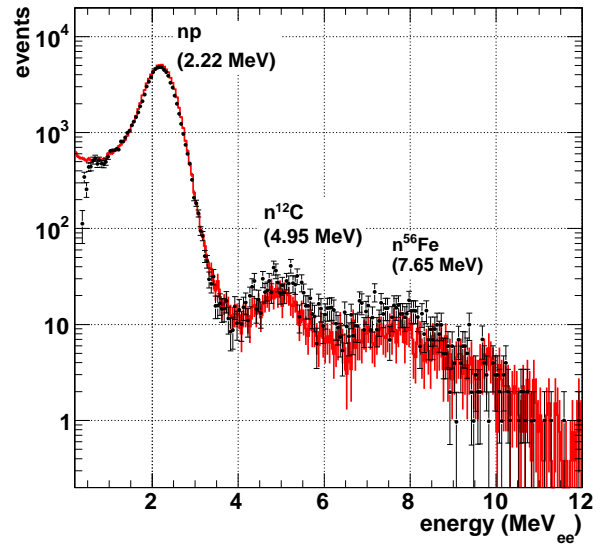}
 \hspace{0.1pc}
 \includegraphics[width=20pc,angle=0]{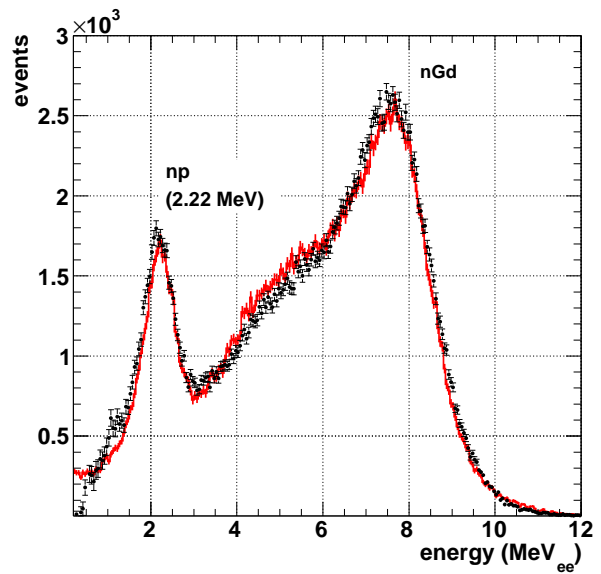} 
 \caption{\label{cap} Energy spectrum of the neutron capture signals (background subtracted) (in black), compared with simulation (in red). Top and bottom panel are relative to LVD-scint and LVD-Gd scintillator respectively.}
 \end{center}
\end{figure}

In the present work we will refer to the light output in terms of the scintillator response to electrons, as this can be taken to be linear (within $\sim$1$\%$ for electrons of energy E $>$ 2 MeV). The measured signals will then be expressed in MeV electron recoils equivalent (MeV$_{ee}$).\\
Assuming that our scintillator has a response function to nuclear recoils consistent with that of similar organic scintillators, we should expect signals up to $\sim$2 MeV$_{ee}$ for neutrons of 6.5 MeV which elastically scatter protons. Neutrons of lower energy are accompanied with a 4.44 MeV gamma emission, according to equations \ref{eqn1} and \ref{eqn2}.\\
Neutron captures on H are followed by the emission of the 2.22 MeV $\gamma$ quantum from the deuterium de-excitation; while captures on Gd originate $\gamma$ cascades of total energy $\sim$8 MeV.
As we want to measure the prompt signals with a threshold as low as possible, we decided to trigger on the delayed pulse and then look backward in time to find the prompt one.
The trigger condition to enable samples storage was chosen to be a 3-fold coincidence of the PMTs signal exceeding a threshold of 1.8 MeV$_{ee}$ for the counter with LVD-scint and 3.2 MeV$_{ee}$ for the counter with LVD-Gd.
The background rate at these thresholds, for a counter placed underground, is about 50$\%$ and 2$\%$ of the neutron rate from the source.\\
Neutron events are selected requiring that the delayed pulse must be accompanied, in a time window of 320 $\mu$s preceding the trigger, by a second signal (prompt) detected with a ZLE threshold of 0.8 MeV$_{ee}$ for both the detectors.
This additional condition helps to get rid of the background enanching the signal to noise ratio of about a factor ten.\\
The highest-energy prompt signals are large enough to trigger by themselves, in this case the delayed pulse will be left out of the acquisition window.
To avoid this situation the acquisition window has been increased to 640 $\mu$s moving the trigger in the middle of it.
In this way we can consider the trigger pulse as due to nuclear recoil or nuclear capture, depending on the presence of a pulse in the following or in the preceding 320 $\mu$s respectively. 
In turn the pulse in the pre-trigger region can be regarded as a recoil signal and used to extend the recoil energy spectrum down to 0.8 MeV$_{ee}$.\\
In fig. \ref{dt} we show the distribution of the absolute value of the difference between the trigger pulse and the pre-trigger or post-trigger pulse if present in the same event, for both the scintillators measured.
Fitting the distribution with an exponential function we found, for LVD-scint, a mean neutron capture time of 210 $\mu$s, well in agreement with previous measurements taking into account that the neutron source was placed in the center of the detector.
Concerning the counter filled up with LVD-Gd, doped at a Gd concentration of 0.93 g/l, we found a mean neutron capture time of 27$\mu$s. Also in this case the measured value is in agreement with previous measurements and simulations \cite{gdperf}.
In the same figure, gray points represent the background level, obtained applying the same analysis to a dataset collected after removing the Am-Be source.\\
We developed a Monte Carlo simulation based on the GEANT4 toolkit \cite{GEANT4} implementing the detector description and the Am-Be source energy spectrum taken from \cite{marsh}.\\
GEANT4 has high precision transport models (G4NeutronHPElastic and G4NeutronHPInelastic) available for neutrons of energies up to 20 MeV and for all materials. The high precision models are based on the Evaluated Nuclear Data Files (ENDF/B-VI) \cite{endf}, which contains tabulated cross-sections for elastic reactions and the different inelastic final states (e.g.: n$\gamma$, np, nd, etc.). The high-quality ENDF/B-VI data generally allows for accurate simulations of these low-energy neutrons \cite{plastic}.\\
The outcome of the simulation is shown in fig. \ref{cap} where the energy spectra of the de-excitation $\gamma$-ray following neutron capture are compared with experimental ones.
The 2.22 MeV gamma emitted after neutron capture on $^{1}$H, clearly visible for both the detectors, has been used to determine the energy scale with an uncertainty of $\pm$1.5$\%$.\\
The spectrum relative to the unloaded scintillator exhibits the peaks due to neutron captures on $^{12}$C (4.95 MeV), and $^{56}$Fe (7.65 MeV), which are overwhelmed in the other detector by the presence of gadolinium.\\
The light output produced by the nuclear recoils was evaluated from the eq. \ref{eqq}, taking into account the contribution of both the ions $^{1}$H and $^{12}$C, even if the latter was rather negligible when compared to the former.\\
The stopping power $dE/dx$ in liquid scintillator, from which the quenching factor depends, was taken from the ESTAR database \cite{nist} for electrons, whereas for protons and heavier ions it was calculated with the SRIM code \cite{srim}.\\
For a specific material, the eq. \ref{eqq} has only one free parameter, $kB$, the so-called Birks factor, which has been determined by fitting, with the least-squares method, the simulated energy spectrum of nuclear recoils to the experimental one.
In the fitting procedure, the comparison between experimental data and simulation was performed by re-weighting Monte Carlo events for different choice of the $kB$ parameter.
\begin{figure}[htb]
\begin{center}
 \includegraphics[width=20pc,angle=0]{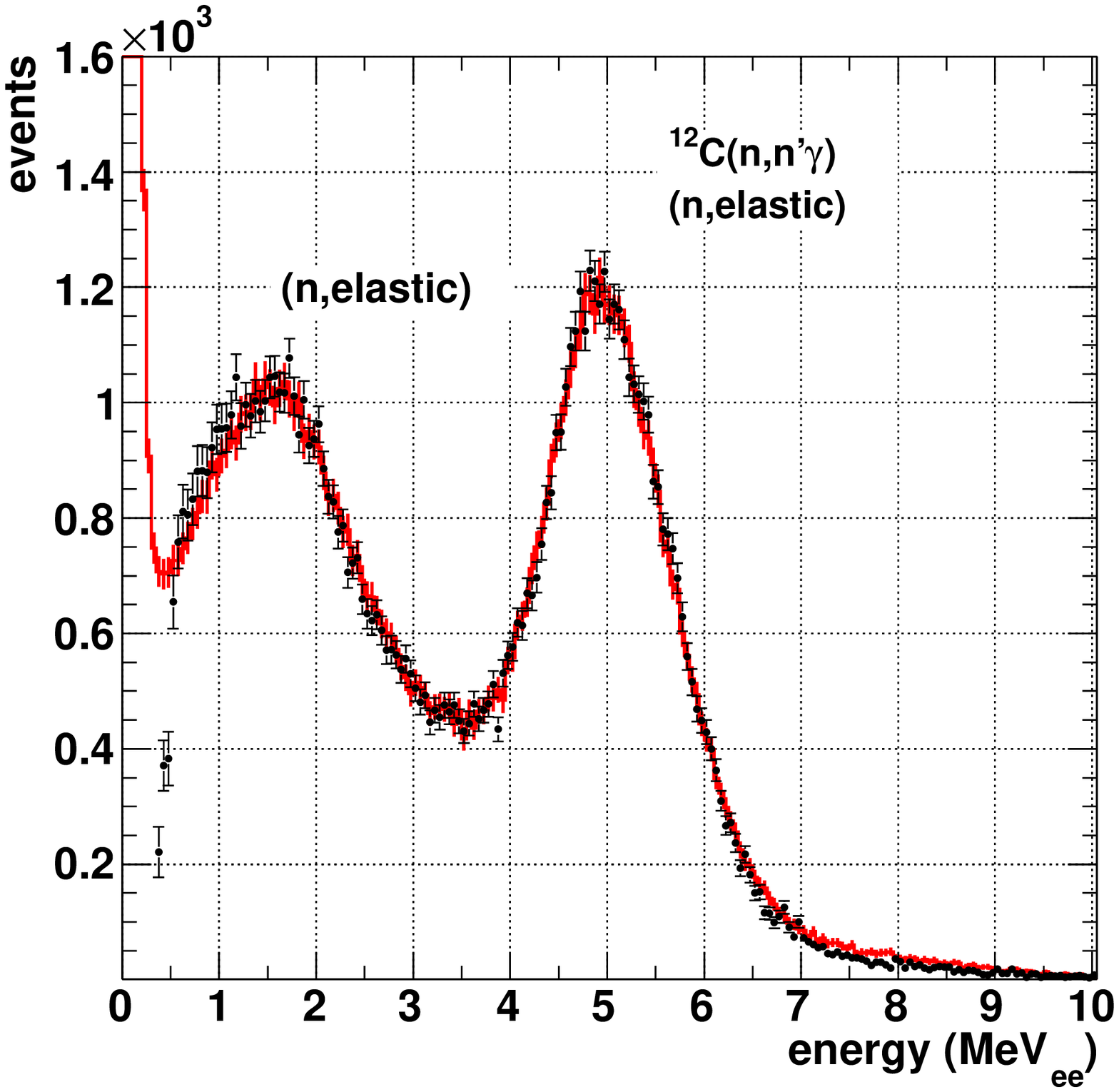}
 \hspace{0.1pc}
 \includegraphics[width=20pc,angle=0]{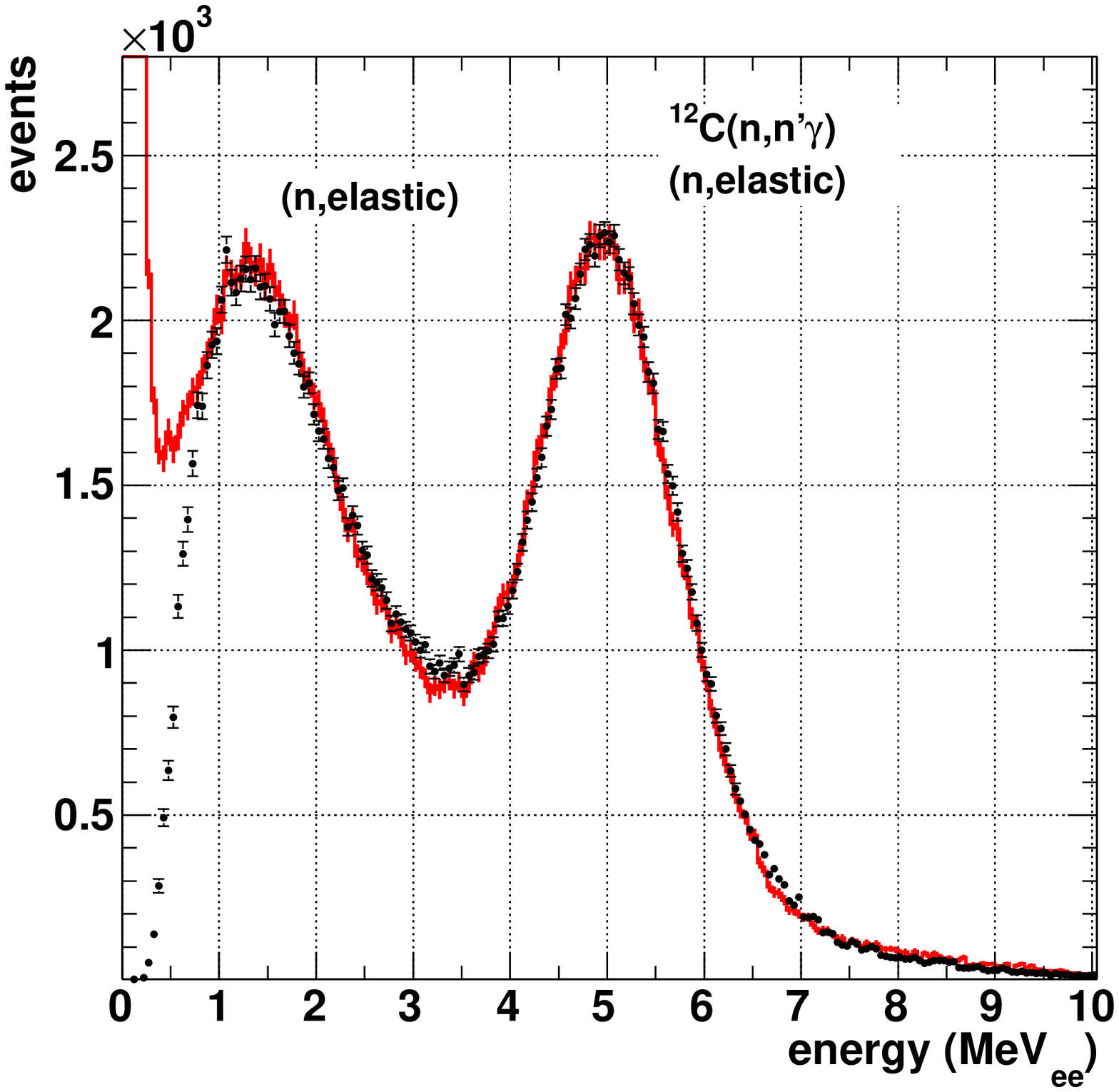} 
 \caption{\label{rec}Energy spectrum of the prompt signals after background subtraction (in black), compared with the simulation (in red). The detection threshold is at about 0.8 MeV$_{ee}$.
 The value of the Birks factor which minimize the $\chi^{2}$ is $kB = 0.0125\; g\; cm^{-2} MeV^{-1}$ for LVD-scint (top panel) and $kB = 0.0140\; g\; cm^{-2} MeV^{-1}$ for LVD-Gd (bottom panel).}
 \end{center}
\end{figure}
The parameter estimation and the goodness of fit are based on the $\chi^2$ test, modified for comparing weighted and unweighted events \cite{weighted}.\\
The result of the fitting procedure is shown in fig. \ref{rec}, where experimental and simulated date are represented with black and red line  respectively.
The two spectra diverge at E$\lesssim$0.8 MeV$_{ee}$ because of the detection threshold. The first peak at E$\simeq$1.5 MeV$_{ee}$ is due to neutron elastic scattering, whereas the second peak is due to the 4.44 MeV gamma from $^{12}$C$(n,n^{\prime}\gamma)$ in addition to elastic scattering of lower energy neutrons (E$_n$ $<$6.5 MeV).\\
The probability to get by chance a $\chi^2$ higher than the observed is 34\% for the Russian scintillator and 17\% for the Italian one.\\
The minimum $\chi^2$ corresponds to a Birks parameter value $kB = 0.0125\; g\; cm^{-2} MeV^{-1}$ and 
$kB = 0.0140\; g\; cm^{-2} MeV^{-1}$ for LVD-scint and LVD-Gd respectively.
\begin{figure}[htb]
\begin{center}
 \includegraphics[width=20pc,angle=0]{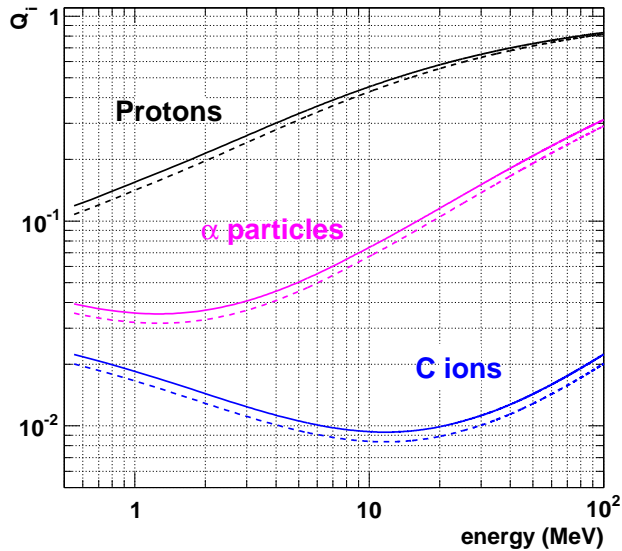}
 \caption{\label{curve} Quenching factors for protons, alpha particles and carbon ions obtained from the eq. \protect\ref{eqq} using the best fit parameter $kB = 0.0125\; g\; cm^{-2} MeV^{-1}$ for LVD-scint (solid lines) and $kB = 0.0140\; g\; cm^{-2} MeV^{-1}$ for LVD-Gd (dashed lines).}
 \end{center}
\end{figure}
By letting $\chi^2$ vary by one unit we found the standard error interval, which is $\pm1\%$ in both cases.
In addition to the statistical error we estimated a systematic contribution of $\pm 5\%$ coming mostly from the uncertainty in the energy calibration.\\
According to the author of the work \cite{semiemp}, the Birks factor is independent of the particle type and can then be used afterwords to calculate quenching factors for other particles and in other energy regions.\\
In fig. \ref{curve} we show the resulting quenching curve for protons, alpha particles, and carbon ions.
\section{Conclusion}
Two different detectors of the LVD experiment have been exposed to an Am-Be ($\alpha$,n) source, with the purpose of measuring the light output response to nuclear recoils.\\
The delayed signal due to de-excitation $\gamma$-ray following neutron capture has been used to tag the events and effectively reject the gamma background.\\
The analysis is based on comparison of the experimental nuclear recoil spectra with Monte Carlo simulation.
We found that the light output response is well described by the Birks model and a Birks factor kB = (0.0125 $\pm$ 0.0004)
$g\; cm^{-2} MeV^{-1}$ and kB = (0.0140 $\pm$ 0.0007) $g\; cm^{-2} MeV^{-1}$ was obtained from the two scintillators LVD-scint and LVD-Gd respectively.\\
Quenching factors for ions in scintillator have then been calculated up to 100 MeV.\\
This result can be of interest for fast and high-energy neutron spectroscopy performed with the LVD detector.
It can be directly applied to 73\% of the whole LVD array (Russian scintillator, i.e. LVD-scint) concerning the remaining amount of counters, filled up with Italian scintillator, we have not yet made a direct measurement of the quenching factors. Nevertheless, we could probably consider the results coming from the Italian Gd-doped scintillator as a lower bound for the undoped one.

\acknowledgments{We are very grateful for helpful discussion with Dr. A.Ianni, Dr. M.Laubenstein and Dr. L.Pandola.}

\end{document}